Magnetotransport study of the mini-Dirac cone in AB-stacked four- to six-layer graphene under perpendicular electric field

Kota Horii[1], Tomoaki Nakasuga[1], Taiki Hirahara[1], Shingo Tajima[1], Ryoya Ebisuoka[1], Kenji Watanabe[2], Takashi Taniguchi[2], and Ryuta Yagi[1]

[1]Graduate School of Advanced Sciences of Matter (AdSM), Hiroshima University, Higashi-Hiroshima 739-8530, Japan.

[2]National Institute for Materials Sciences (NIMS), Tsukuba 305-0044, Japan.

ABSTRACT

Landau levels of AB-stacked four- to six- layer graphene were studied experimentally in the presence of a perpendicular electric field. The Landau levels at a low magnetic field showed a characteristic structure that originated from the mini-Dirac cones created by the perpendicular electric fields. Six-fold or twelve-fold degenerate Landau levels arising from the mini-Dirac cones were observed in the vicinity of the charge neutrality point as a radial structure in a plot of their Landau fan diagrams. The structure of four-layer graphene had approximate electron-hole symmetry near the charge neutrality point, while the five- and six-layer graphene showed asymmetry. Numerical calculations of the dispersion relation and Landau level spectra indicated that trigonal warping played an essential role in forming the experimentally observed Landau level structure in low magnetic fields. In high magnetic fields, trigonal warping becomes less significant and the Landau level spectra become simpler, approaching those that were calculated without considering trigonal warping. The appearance of mini-Dirac cones in a perpendicular electric field would be a common feature of AB-stacked multilayer graphene.

(*Corresponding author: yagi@hiroshima-u.ac.jp)

## 1. Introduction

Monolayer graphene has a Dirac cones with a massless dispersion relation [1-3]. Because the velocity of its electrons is expected to be about 1/300th the speed of light, much faster than electrons in conventional semiconductors even at low carrier densities, graphene is expected to be a candidate materials for next-generation devices [4-5]. Graphene also has a unique property of controllability of its band structure, *i.e.*, stacking the graphene layers dramatically changes the band structure [6-34]. The low energy band structure of bilayer graphene is approximately parabolic, and the electrons in it can be regarded as massive Dirac fermions [1,6-8,26]. When there are more than three layers, two-different stackings, *i.e.*, ABA and ABC, are possible in terms of the relative positions of the carbon atoms which form dimers in adjacent layers. In AB-stacked graphene (*i.e.*, the ABA stacking), the band structure systematically evolves as the number of layers increases. $2N$ ($N$:integer) layer graphene has $N$ sets of bilayer-like bands, whereas $2N + 1$ layer graphene has an extra mono-layer like band in addition to the $N$ sets of bilayer-like bands [14-15,17,26,32,35-37].

The even-odd effect implies that AB-stacked graphene with an even number of layers does not have Dirac cones. However, recent studies have indicated that Dirac cones are created by applying a perpendicular electric field. For example, Morimoto and Koshino [38] theoretically showed that a perpendicular electric field forms an energy gap between the bottoms of the bilayer-like bands, but the gap closes locally as a result of trigonal warping, and thereby mini-Dirac cones form.

Mini-Dirac cones are expected to appear significantly in AB-stacked graphene of many layers. The recently found intrinsic resistance peak structures arising from topological changes in the Fermi surface [27,29-30] are expected to be partly due to mini- Dirac cones. In AB-stacked four-layer [27,29,31], five-layer [31], six-layer [30], and seven-layer graphene [31], the carrier density dependence of the resistivity shows peak structures specific to the band structures, and maps of the resistivity as a function of perpendicular electric flux density and carrier density show ridge structures. [27,29-31]. Numerical calculations of the dispersion relation in the presence of a perpendicular electric flux density indicated that the conspicuous ridges are due to mini-Dirac cones formed by the field.

A previous study examined the Landau level structure in AB-stacked tetralayers in the presence of perpendicular electric field in relation to the appearance of mini-Dirac cones [27]. Six-fold degenerate Landau levels associated with the mini-Dirac cones are reported. In the current study, we performed a systematic and detailed study on the Landau level structure of AB- stacked four- to six-layer graphene in a perpendicular electric field. We observed that, in a perpendicular electric field, the mini-Dirac appeared in AB-stacked graphene having several layers, and we interpreted the observed Landau level spectra in terms of the shapes of the dispersion relations.

Here, we briefly describe the variation in the dispersion relations arising when a perpendicular electric field is applied. Figure 2 shows plots of numerical calculation for AB-stacked four- to six-layer graphene for $D_\perp = 0$ (top panels) and $2.4 \times 10^{-7}$ cm$^{-2}$As (bottom panels). The calculations were based on the effective mass approximation with all the Slonczewski-Weiss-McClure parameters of graphite. The dispersion relations for $D_\perp = 0$ showed systematic variations, as indicated by the even-odd effect; a linear band can be seen only in the five-layer graphene [12, 9, 17, 25, 27-37,47-49], whereas the band structure is rather more complicated near $E = 0$ [26,36,48] than the simple expectation that mono- and bilayer-like bands have linear and parabolic dispersions [9,26,35,37]. In a perpendicular electric field, the dispersion relations become significantly different from those for $D_\perp = 0$. The field opens energy gaps between the bottoms of the bilayer-like bands, as in the case of bilayer graphene; however, the energy gap tends to close locally and mini-Dirac cone structures appear (left arrows), which have linear dispersion relations over a limited ranges of k near $E = 0$. In addition, for the six-layer case, a few mini-Dirac cones formed at energies away from $E = 0$. One can also recognize cone-like structures exhibiting energy gaps at their vertices, as indicated by the right arrows.

The mini-Dirac cones originate from the trigonal warping inherent to graphene. Trigonal warping locally closes the energy gaps of the bilayer-like bands that are created by a perpendicular electric field [38]. This results in a change in the band mass as well as the shape of the Fermi surface, and induces a topological change of the Fermi surface when the chemical potential and perpendicular electric fields are varied by changing the gate voltages [27,29-31].

2. Experimental

Graphene was prepared by mechanically exfoliating high-quality Kish graphite crystal with adhesive tape [1]. The number of layers and stacking were determined by combining information from different methods of measurement. The number of layers in the graphene flakes was roughly determined from the color signal intensity of the digitized image. The shape of the Raman D' band spectra was measured for many graphene flakes for which the number of layers was determined by atomic force microscopy (AFM) topography. The number of layers and the stacking of graphene samples were determined by referring the spectra shapes [32,39-43]. We also checked the number of layers and stacking via the Landau level structure [27-28,32].

Figures 1(a) and 1(b) show an optical micrograph and the schematic structure of a typical graphene sample used in the present study. The samples consisted of *h*-BN encapsulated graphene equipped with top and bottom gate electrodes. Each sample was encapsulated with a transfer technique utilizing organic polymers, poly-propylene carbonate (PPC), and dimethilpolysiloxane (PDMS) [46]. The bottom gate electrode was a P[++] doped Si substrate covered with SiO$_2$ (300nm thick). The top gate electrode was formed by placing thick graphene on the encapsulated graphene stack. The sample geometry was defined by reactive ion etching using a mixture of low pressure CF$_4$ and O$_2$ gas. The geometry of the effective sample area was identical to that of the top gate electrode. Electrical contact with the graphene was attained by using the edge-contact technique [46].

The top and bottom gate electrodes were used to control the carrier density and perpendicular electric field independently. The carrier density induced in the graphene is thus given by the sum of the carriers induced by the gate voltages:

$$n_{tot} = C_t(V_t - V_{t0}) + C_b(V_b - V_{b0}), \tag{1}$$

where $C_t$ and $C_b$ are the specific capacitance of the top and bottom gate electrodes and, $V_{t0}$ and $V_{b0}$ are for the offset voltages associated with the gate electrode The difference in the charge density induced by the two electrode is proportional to the perpendicular electric flux density $D_\perp$, which is given by

$$D_\perp = \left(\frac{e}{2}\right)\left(C_t(V_t - V_{t0}) - C_b(V_b - V_{b0})\right). \tag{2}$$

Here, $C_b$ is about 108 aF/μm, and $C_t/C_b$ varies depending on thickness of top gate insulator (*h*-BN); it typically ranged from 3 to 5.

## 3. Results and Discussion

### 3.1 AB-stacked four-layer graphene

Although the Landau level structure for AB-stacked four-layer graphene has been reported in Ref. [27], we briefly show results not only for providing complementary information, but also illustrating the systematic evolution with respect to the number of layers. Figure 3 (a) shows a map of resistivity as a function of the top and bottom gate voltages, measured at zero magnetic field ($B = 0$ T) and at temperature of 4.2 K. Large resistivity peaks appearing linearly as a function of $V_t$ and $V_b$ occurred at the charge neutrality condition ($n_{tot} = 0$) where the carrier densities induced by the top and bottom gate voltages are compensated. From this relation, the specific capacitance ratio was estimated to be $C_{tg}/C_{bg} \approx 4.6$. The electrical mobility of this sample was estimated to be above $4 \times 10^4$ cm$^2$/Vs in the large carrier density regime (bottom panel in Fig. 1). At larger $n_{tot}$ values, one can recognize ridges of resistivity peaks, as reported in Refs. [27,29]. Fig. 3(b) is a replot of resistivity against $n_{tot}$ and $D_\perp$, which shows conspicuous resistance ridge structures specific to the AB-stacked four-layer graphene [27,29]

The formation of mini-Dirac cones was verified by making magnetotransport measurements at $T = 4.2$ K. The longitudinal magnetoresistance ($\rho_{xx}$) showed Shubnikov-de Haas oscillations, which were complicated because of beating due to different bands. Figure 4 (a) is a map of $\rho_{xx}$ as a function of carrier density $n_{tot}$ and magnetic field for $D_\perp = 3.8 \times 10^{-7}$ cm$^{-2}$As. Figure 4(b) shows a map of $d\rho_{xx}/dB$, which allows one to observe small variations in magnetoresistance. The Landau in the presence of the perpendicular electric fields shows a significant variation in shape from those at zero perpendicular electric field (Figs. 4(d) and 4(e)).

Next, let us examine the difference between the Landau level structures in the absence and presence of a perpendicular electric field. For $D_\perp = 0$, the zero-mode Landau levels are discernible as vertically elongated beating structures at $n_{tot} \sim -0.2,$ and $0.8 \times 10^{12}$ cm$^{-2}$ (shown by the arrows in Fig. 4(d)); these are approximately reproduced by the numerically calculated density of states (Fig 4(f)). It can be seen that the level structure near $|n_{tot}| < 1 \times 10^{12}$ cm$^{-2}$ is asymmetric between the electron and the hole regime. The above-mentioned structure changes drastically in the perpendicular electric field. In Figs. 4(a) and 4(b), two vertically elongated Landau levels ($\alpha$, $\beta_e$, $\beta_h$), and parabolic structures ($\gamma_e$ and $\gamma_h$) can be seen in the vicinity of $n_{tot} = 0$. Unlike the results for $D_\perp = 0$, these Landau level structures are approximately symmetric between the electron and hole regimes. This indicates that mini-Dirac cones form near the charge neutrality point, and the dispersion relations near $E = 0$ became nearly symmetric between the electron and hole regimes.

More evidence of mini-Dirac cones can be found by examining the numerically calculated density of states in a magnetic field (Fig. 4(c)); as indicated, the observed Landau level structure is approximately reproduced. The density of states is based on the results of a calculation that considers the electrostatic potential of each layer, which results from screening of external electric fields induced by the top and bottom gate voltages. We assumed that the carriers induced by each gate voltage attenuate exponentially with distance, with a characteristic screening length of $\lambda = 0.45$ nm. The experimental Landau level α in Fig. 4(b), which was 12-fold degenerate, corresponds to levels $\alpha_h$ and $\alpha_e$ in Fig. 4(c), although an energy gap between $\alpha_h$ and $\alpha_e$ cannot be clearly seen, possibly because it occurred at the charge neutrality point where the resistance showed a large peak. On the other hand, Landau levels $\beta_e$ and $\beta_h$ in Fig. 4(c) are both six-fold degenerate and can be seen in Fig. 4(a) and 4(b). The Landau level structures $\gamma_h$ and $\gamma_e$ (in Fig. 4(b)), which are parabolic in shape, can also be seen in the calculated density of states.

The characteristic structures of the density of states result from the Landau level spectra. The upper and lower panels in Fig. 5(a) show the dispersion relation and Landau level spectra in a perpendicular electric field. The results for zero-electric field are displayed in Fig. 5(b) for comparison. In Fig. 5(a), Landau levels $\alpha_e$, $\alpha_h$, $\beta_e$, and $\beta_h$ are conspicuous near $E = 0$ meV. Each of these levels consists of three levels, which are entangled and form a bundle [27]. They seem to be degenerate at sufficiently low magnetic field and split weakly with increasing magnetic field. The origin of the Landau level structure can be qualitatively understood by comparing the spectra with the dispersion relation (the upper panel of Fig. 5(a)). Two sets of mini-Dirac cones form at each of K and K' point. The one cone is larger than the other, which is formed at about $\hbar k_x v_0/\gamma_1 \approx 0.4$ at the K point ($-0.4$ at the K' point). The vertices of the cones are approximately at $E \approx 0$ (labeled by $a$), and no energy gap appear near the vertices as similar to the case of mono-layer graphene. The other mini-Dirac cone appears at $\hbar k_x v_0/\gamma_1 \approx -0.2$ at the K point (0.2 at the K' point). The peak height of the smaller cone is considerably lower than that of the larger one because an energy gap opens between the vertices (labeled $b_e$ and $b_h$). There are three identical larger (smaller) mini-Dirac cones because of the three-fold rotational symmetry arising from trigonal warping (for example, the case of the large mini-Dirac cones is schematically illustrated in Fig. 5(c)).

The energies of Landau levels $\alpha_e, \alpha_h, \beta_e,$ and $\beta_h$ extrapolated to zero magnetic field are approximately the same as the energies for the vertices of the mini-Dirac cones (Fig. 5(a)). As $B \to 0$, the bundle of Landau levels $\alpha_h$ and $\alpha_h$ tends to $E \approx 0$ meV, which is the same energy as point $a$ in the dispersion relation. Considering that the variation in energy of these levels with respect to the magnetic field is rather smaller than other Landau levels, $\alpha_h$ and $\alpha_h$ must be zero-mode Landau levels for the larger mini-Dirac cones. Similarly, the bundles of Landau levels $\beta_h$ and $\beta_e$ appear approximately at the same energy as the vertices $b_h$ and $b_e$ of the smaller mini-Dirac cones. Therefore, these levels would also correspond to the zero-mode Landau levels of the mini-Dirac cones. The Landau levels of the cones with higher Landau indices are also visible. For $|E| < 10$ meV and $B < 0.5$ T, it can be seen that the energy of the Landau levels is nearly proportional to the square root of the magnetic field, as schematically shown in Fig. 5 (c). The square root dependence is a specific feature of the Landau levels arising from massless-Dirac fermions, *e.g.*, in monolayer graphene. These levels would have originated from the larger mini-Dirac cones. The structure is hardly visible at large $|E|$, because other Landau levels are overlaid. However, Landau levels with index $|N| = 1$, are visible in the experimental fan diagram and in the numerically calculated density of states ($\gamma_e$ and $\gamma_h$ in Fig. 4(b)).

The bundles of Landau levels $\alpha_h$, $\alpha_h$, $\beta_h$, and $\beta_e$ each consist of three Landau levels, so they are six-fold degenerate in the limit $B \to 0$ if spin degeneracy is considered. The three-fold degeneracy, except for the spin degeneracy, would originate from the three-fold rotational symmetry of the mini-Dirac cones. The lifting of the degenerate Landau levels and formation of entangled bundles with increasing magnetic field are possibly due to mixing of the levels between the mini-Dirac cones. At this magnitude of $D_\perp$, the Dirac points of the larger mini-Dirac cones are separated by an energy barrier of about 15 meV in *k*-space. If the energy barrier is sufficiently large, the Landau levels of each cone would form locally in each mini-valley at sufficiently low magnetic fields. In the opposite case, *i.e.*, in the case of a smaller energy barrier and a large magnetic field, the coupling of the Landau levels between the valley (Fig. 5(c)) would become increasingly large, and thereby, lifting of the degeneracy would become significant. At much higher magnetic fields, the Landau level structure becomes simpler, appearing as a superposition of Landau levels of the two sets of gapped bilayer bands. This is reminiscent of magnetic breakdown in the semi-classical theory of the electronic band structure at high magnetic fields [50-54], as will be described in detail in Sec. 4.

### 3.2 Mini-Dirac cones in AB-stacked five-layer graphene

Characteristic Landau levels arising from mini-Dirac cones also appeared in the measurement of AB-stacked five-layer graphene. Figs. 6(a) and (b) show maps of $\rho_{xx}$ and $d\rho_{xx}/dB$ measured for $D_\perp = 1.72 \times 10^{-7}$ cm$^{-2}$As. The Landau-level structure, in particular, near $n_{tot} = 0$, is strikingly different from that for $D_\perp = 0$ (Figs. 6(d), 6(e)), which can be explained in terms of two sets of bilayer-like bands and a monolayer-like band [31-32]. The Landau levels for $D_\perp \neq 0$ exhibit characteristic structure that extends radially from $n_{tot} = 0$ and $B = 0$, as in the case of the four-layer graphene. However, unlike the four-layer case, the conspicuous energy gaps (or minima of the Shubnikov-de Haas oscillations) have electron-hole asymmetry; they are discernible at $\nu = 12$ and $-14$ in the map of $d\rho_{xx}/dB$. One can also see minima around $\nu = -26$. The Landau levels between $\nu = 12$ and $-14$ have a complicated structure, possibly consisting of multiple levels that cross each other (Fig. 6(b)). Similarly, the electronic states between $\nu = -26$ and -14 consist of multiple Landau levels. The observed structure can be approximately explained by the numerical calculation of the Landau level spectra and their density of states. First, let us compare the experimental results with a map of the density of states (Fig. 6(c)). The asymmetric structure of energy gaps ($\nu = 12$ and $-14$) is reproduced in Fig. 6(c). The electronic states between the gaps consist of multiple Landau levels. These energy gaps do not exist in the case of $D_\perp = 0$ (Fig. 6(e)); the Landau levels approximately consist of a superposition of two sets of fans arising from bilayer-like bands with a split zero-mode (as indicated by the arrows in Fig. 6(c)).

The relation between the Landau level spectra and the dispersion relation was studied in order to get a qualitative understanding of the Landau levels at low magnetic fields. Figs. 7 (a) and (b) show the dispersion relations and the Landau level spectra, for $D_\perp = 1.72 \times 10^{-7}$ and 0 cm$^{-2}$As, respectively. In Fig. 7(a), large energy gaps with filling factors, $\nu = -14$, -2, and 12 form near the charge neutrality point between the entangled Landau levels forming bundles, Each of the bundles consists of multiple Landau levels. The bundles converge to single levels as $B \rightarrow 0$T. As shown in the lower panel of Fig. 7 (a), there are vertices of mini-Dirac cone structures ($b_e$, $b_h$, and $a$) at those energies. Two bundles with three-fold degenerate Landau levels (except for spin) originate from $a$, which is a vertex of a mini-Dirac cone. The vertices $b_h$ and $b_e$ are those of a mini-Dirac cone-like structure which has an energy gap between the vertices. A bundle with three-fold degenerate Landau levels (except for spin) stems from each of $b_e$ and $b_h$. These three-fold degenerate Landau levels can be interpreted as zero-mode Landau levels associated with the vertices of the mini-Dirac cones, as in the case of four-layer graphene.

Besides the above bundles, there is a nondegenerate (except for spin) zero-mode Landau level associated with vertex $m$, which is located at $k_x = 0$. This principally originates from the monolayer-like band, which forms sharp peak structures at $k_x = 0$ for $D_\perp = 0$. In the dispersion relation shown in Fig. 7(a), the mini-Dirac cone structure of the monolayer-like band appears only in the hole-like band. One cannot clearly see a similar structure in the electron-like band. However, a zero-mode Landau level corresponding to an electron-like band is present at $E \approx$ 13 meV.

The presence of the monolayer band significantly affected the filling factors of the large energy gaps, as compared with the four-layer case: the energy gaps occurred at $\nu = -14$ and 12 in the AB-stacked five-layer graphene, while they appeared at $\nu = -12, -6, 6$ and 12 in the four-layer sample. The shift of $-2$ in the filling factor is due to the presence of the zero-mode Landau level associated with vertex $m$. This reflects the even-odd layer number effect in the band structure of AB stacked graphene [14-15,17,26,32,35-37].

One can recognize the Landau levels of mini-Dirac cones with higher indices in the Landau level spectra. For example, a hole-like Landau level with index 1 is discernible as the bundle labeled $\alpha$. This bundle consists of six Landau levels roughly between $-26 < \nu < -14$ in the experiment (Figs. 6 (a) and 6(b)). An electron-like Landau level with index one is also discernible as $\beta$ in Fig. 7(a). However an energy gap associated with the bundle is expected to appear only at low magnetic fields ($B < 0.5$T), so it would have been barely visible in the experiment.

A correspondence between the zero-mode Landau levels and the vertices of the mini-Dirac cones can also be found in the case of $D_\perp = 0$. The monolayer-like band forms gapped Dirac cones with vertices $m_h$ and $m_e$ (Fig. 7(b)). The energies correspond to those of non-degenerate (except for spin) zero-mode Landau levels arising from the monolayer-like band. As for the bilayer-like bands, the zero-mode Landau levels approximately stem from the bottoms.

### 3.3 Mini-Dirac cones in AB-stacked six-layer graphene

Although the band structure of the six-layer sample was much more complicated than the above-mentioned results, characteristic mini-Dirac cones structures still appeared in the Landau fan diagrams. Figures 8(a) and 8(b) show maps of $\rho_{xx}$ as a function of $n_{tot}$ and $B$, for $D_\perp = 1.42 \times 10^{-7}$ and $0$ cm$^{-2}$As, respectively. Figures 8(c), and (d) show maps of the numerically calculated density of states for the same values of $D_\perp$, and they approximately reproduce the observed maps of $\rho_{xx}$. The Landau levels in a perpendicular electric field show an approximately radial structure with a center located at $n_{tot} = 0$ cm$^{-2}$ and $B = 0$ in the vicinity of the charge neutrality point, as in four- and five-layer cases. One cannot recognize any such structure at $D_\perp = 0$. The numerically calculated dispersion relation (lower panel in Fig. 9(a)) revealed that there are two sets of mini-Dirac cones near $n_{tot} = 0$, i.e., $|E| < 10$ meV; one has vertices $b$ and $c$; the other one has vertices $a$ and $e$. The electron-like mini-Dirac cone with vertex $c$ and the hole-like mini-Dirac cone with vertex $a$ overlap and form a semi-metallic band structure with a small compensated carrier density. The superposition of the Landau levels associated with these mini-Dirac cones form complicated spectra shapes. Nonetheless, the characteristic V-shaped structure is visible in the numerically calculated density of states as well as the experimental maps of $\rho_{xx}$ (marked by A in Fig. 9(b)).

What is particular to the six-layer graphene as compared with samples with fewer layers is that the mini-Dirac cones are expected to form not only in the vicinity of $n_{tot} = 0$, but also at larger values of $|n_{tot}|$ [30]. The mini-Dirac cones with vertices $f, g$ and $h$ are in the electron regime, while those in the hole regime are smaller ones with vertices $i$ and $j$. The mini-Dirac cones in the electron regime form V-shaped structure in the map of resistivity around $n_{tot} = 1.5 \times 10^{12}$ cm$^{-2}$ (marked by A in Fig. 9(b)), which is similar to the structure formed in the vicinity of $n_{tot} = 0$ in the AB-stacked four-layer graphene (Figs. 4(a)-4(c)). This should be because the structure of the dispersion relation between $E \approx 5$ and $40$ meV (lower panel of Fig. 9(a)) are qualitatively the same as that of the four-layer graphene except for the presence of an extra bilayer-like band with larger wave numbers. In both six- and four-layer graphene, there are a gapless and a gapped mini-Dirac cone (the vertices of $f, g$ and $h$ in Fig. 9(a) correspond to $a$, $b_h$, and $b_e$ in Fig. 5(a)). Near the energy of these vertices, zero-mode Landau levels of mini-Dirac cones appear as vertically elongated levels at low magnetic fields (marked with B in Fig 9(b)). As for the hole regime, one such structure associated vertices $i$ and $j$ is visible in the experimental result as well as in the numerical simulation (marked C in Fig 9(b)). This means that the shape of the dispersion relation between $E \approx -30$ and $-10$ meV significantly differs from the four-layer case.

4. Discussion

Trigonal warping plays an essential role in forming mini-Dirac cones in AB-stacked multilayer layer graphene in a perpendicular electric field. If we assume no trigonal warping, rotationally symmetric energy gaps form at energies where the bilayer-like bands cross. As a result of the difference in the band structure, the numerically calculated density of states in a magnetic field shows significantly different patterns between cases with and without trigonal warping (see Figs. 10(a)-10(c) for AB-stacked four- to six-layer graphene with $\gamma_3 = 0$ and 0.3 eV). The V-shaped structures associated with mini-Dirac cones are absent for the case with $\gamma_3 = 0$ eV. Thus, the experimentally observed fan diagrams could not be reproduced without trigonal warping.

By comparing the results of the numerical calculation of the dispersion relations and the Landau level spectra, we have empirically shown that the Landau levels associated with the mini-Dirac cones appear at low magnetic fields and near the carrier densities that correspond to the energies of their vertices. An analytic proof for this occurrence of the Landau levels associated with the mini-Dirac cone might require a sophisticated theoretical treatment or complicated calculation. However, one can still understand it qualitatively by considering the similarity to the case of monolayer graphene: the Dirac points for two valleys, *i.e.*, the K and K' points, are separated in *k*-space by electronic states with considerably high energies. The low energy band structure can be well approximated by a linear dispersion relation calculated for each valley. As for the mini-Dirac cones, the relevant energy is much smaller than the monolayer case, so that the electronic band structure could be calculated for each single mini-Dirac cone for energies in the vicinity of the mini-Dirac point and at a low magnetic field.

Finally, we point out that the Landau-level structures arising from the mini-Dirac cones tend to disappear at high magnetic fields. Here, we only show the case of AB-stacked six-layer graphene as an example. Figure 11 shows the numerically calculated Landau level spectra up to $B = 14$ T. Results for two cases, one with trigonal warping ($\gamma_3$=0.3 eV), and the other without it ($\gamma_3 = 0$ eV), are shown. The former is identical to the one in Fig. 9(a) for $B < 2.5$ T, and therefore, characteristic Landau level structures due to mini- Dirac cones are present at low magnetic fields. The latter has no mini-Dirac cone structures in the band structure. For large magnetic fields (*e.g*, $B > 10$ T), the Landau levels have approximately the same structures in both cases. The six circled sets of vertically elongated Landau levels would be zero-mode Landau levels that corresponds to the bottoms of the bilayer-like bands. Therefore, mini-Dirac cones disappear at sufficiently high magnetic fields. In the semi-classical picture, the disappearance would be interpreted as formation of new quantum states as a result of a tunneling through energy gaps due to high magnetic fields.

Modification of graphene's band structure has been one of the most interesting topics in the graphene research. Graphene itself does not have band gaps, and to overcome this disadvantage for practical applications, various methods of forming a band gap have been proposed for mono- or bilayer graphene, *e.g.*, making narrow wires or antidot structures [55-62] or applying perpendicular electric fields to bilayer graphene [63-69]. On the other hand, for AB-stacked graphene with numerous layers, the bands become semi-metallic [12,15, 17, 25, 32, 35-36, 48, 70], which is not a preferred property for electronic devices. As has been revealed by the present work, making mini-Dirac cones in multilayer graphene via a perpendicular electric field might lead to a way of controlling the band structure of multilayer graphene for the practical application.

## 5. Summary

Landau level in AB-stacked graphene with four to six layers were studied in terms of the formation of mini-Dirac cones in a perpendicular electric field. The samples commonly showed a mini-Dirac cone structure in a perpendicular electric field, and the zero-mode Landau levels of the mini-Dirac cones formed at the energies of the vertices of the mini-Dirac-cones at low magnetic fields. Levels with larger Landau indices were also observed. Trigonal warping of the band structure was essential to forming the mini-Dirac cone structure in the Landau levels at low magnetic fields. At high magnetic fields, the Landau levels arising from the mini-Dirac cones tended to disappear and changed into ones having simpler structures that were qualitatively the same as those without trigonal warping.

## Acknowledgements

This work was supported by KAKENHI No.25107003 from MEXT Japan.

Appendix

A Calculation of the dispersion relation and Landau levels

The dispersion relations were calculated by using a Hamiltonian based on the tight-binding model using all the Slonczewski-Weiss-McClure parameters [12,15,26,36,48]. The wave functions were expanded in plane waves, and the eigenvalues were calculated numerically. In order to consider the effect of the perpendicular electric field, we calculated potential distributions associated with the carrier distributions induced by the screening of the carriers, which were in turn induced by top and bottom gate voltages, by using the methods described in Refs. [30-32]. Carrier screening has been studied since before the advent of graphene [71-77]. Visscher *et al.* calculated screening effect by using Thomas-Fermi approximations. Guinea studied it within the framework of the random phase approximation. Koshino performed self-consistent calculation of the band structures and distribution of carriers induced by external electric fields [74]. Theoretical studies [71-74] and experiments [75-77] estimate the screening length to be less than a few layers of graphite. We have used model distributions of carrier screening where carriers are induced by application of a gate voltage: for each gate electrode, the induced carriers decay exponentially with a relaxation length of $\lambda$ in the direction perpendicular to the two-dimensional plane. The dielectric constant, $\epsilon/\epsilon_0 = 2$, was used to calculate the electrostatic potential. The electrostatic potential for each layer was added to the diagonal elements of the Hamiltonian. The Landau levels were calculated following Ref. [11]. The wave functions were expanded in Landau functions, and the energy eigenvalues were numerically calculated. The screening length is a fitting parameters in our calculations. In the previous paper reporting intrinsic resistance peak (ridge) structures in AB-stacked multilayer graphene, we chose 0.45 nm [31]. In this work, we used the same value of $\lambda$ for calculating the dispersion relations. This value is approximately the same as the one calculated by Koshino *et al.* (about one layer) [74].

Figure Captions

**Fig. 1** Schematic structure and electrical mobility of graphene sample.
(Top left) Optical micrograph of an *h*-BN encapsulated AB-stacked four-layer graphene sample having top and bottom gate electrodes. A micrograph before connecting electrical leads is shown. I+ and I- stand for current probes. V+ and V- stand for voltage probes. TG stands for top gate. (Top right) Schematic vertical structure of the sample. G is graphene, Au is gold, Si is silicon, hBN is *h*-BN and SiO2 is silicon dioxide. The top gate was carefully formed so as not to make direct contact with graphene. (c) Carrier density dependence of the mobility $\mu = 1/(n_{tot}e\rho)$, where $n_{tot}$ is carrier density, $\rho$ is resistivity and $e$ is electron charge. $T = 4.2$ K.

**Fig. 2** Numerically calculated dispersion relations of AB-stacked four- to six-layer graphene.
Upper and lower panels are results for $D_\perp = 0$ and $2.4 \times 10^{-7}$ cm$^{-2}$As, respectively. From left to right, results for AB-stacked graphene with four to six layers are shown. Calculations were based on the effective mass approximation. The Slonczewski-Weiss-McClure parameters of graphite were used ($\gamma_0 = 3.16$ eV, $\gamma_1 = 0.39$ eV, $\gamma_2 = -0.02$ eV, $\gamma_3 = 0.3$ eV, $\gamma_4 = 0.044$ eV $\gamma_5 = 0.038$ eV, and $\Delta_p = 0.037$ eV). Some mini-Dirac cones are indicated by arrows. Results for K point are shown.

**Fig. 3** Maps of resistivity in AB-stacked four-layer graphene
(a) Map of $\rho_{xx}$ as a function of $V_t$ and $V_b$. $T = 4.2$ K. $B = 0$ T. (b) Replot against $n_{tot}$ and $D_\perp$.

**Fig. 4** Landau fan diagrams of AB-stacked four-layer graphene.
(a) Map of $\rho_{xx}$ as a function of $n_{tot}$ and $B$. $|D_\perp| = 3.8. \times 10^{-7}$ cm$^2$As. $T = 4.2$ K. (b) Replotted as $d\rho_{xx}/dB$. Conspicuous Landau levels are labeled $\alpha$, $\beta_e$, $\beta_h$, $\gamma_e$, and $\gamma_h$. (c) Numerically calculated density of states for the same value of $D_\perp$. The Slonczewski-Weiss-McClure parameters of graphite were used. Some Landau levels are indicated by $\alpha_e, \alpha_h$, $\beta_e$, $\beta_h$, $\gamma_e$, and $\gamma_h$ to show the correspondence with the experimental results. (d) Map of $\rho_{xx}$ for $D_\perp = 0$ cm$^{-2}$As. $T = 4.2$ K. (e) Replotted as $d\rho_{xx}/dB$. (f) Numerically calculated density of states for $D_\perp = 0$ cm$^{-2}$As. Arrows in panels (d) and (f) indicate approximate positions of zero-mode Landau levels.

**Fig. 5** Comparison of Landau level spectra with dispersion relation in AB-stacked four-

layer graphene.

(a) (Top) Numerically calculated dispersion relations for $D_\perp = 3.8.\times 10^{-7}$ cm$^2$As. $v_0 = \sqrt{3}a\gamma_0/2\hbar$, where $a$ is graphene's lattice constant in the out-of-plane direction, and $\gamma_0$ and $\gamma_1$ are the SWMcC parameters. (Bottom) Numerically calculated Landau level spectra. Some points in the dispersion relations are labeled $a$, $b_e$ and $b_h$. The energy for $a$ is approximately the same as those the bundles of the zero-mode Landau levels $\alpha_e$ and $\alpha_h$ supposing that energies are extrapolated to $B = 0$. Similarly $b_e$ and $b_h$ correspond to $\beta_e$ and $\beta_h$, respectively. (b) Similar plots for $D_\perp = 0$. (c) Schematic illustration of mini-Dirac cones with three-fold rotational symmetry. Arrows indicate coupling between the states. (d) Schematic illustration of the energies of Landau levels in mono-layer graphene.

**Fig. 6** Landau fan diagram of AB-stacked five-layer graphene in perpendicular electric field.

(a) Map of $\rho_{xx}$ as a function of $n_{tot}$ and $B$. $D_\perp = 1.72 \times 10^{-7}$ cm$^2$As. $T = 4.2$ K. Positions of some filling factors are indicated by numbers and arrows. (b) Replotted as $d\rho_{xx}/dB$. (c) Numerically calculated density of states. The Slonczewski-Weiss-McClure parameters of graphite were used. Conspicuous energy gaps for $\nu = -14$ and 12 are indicated by arrows. (d) Map of $\rho_{xx}$ for $D_\perp = 0$ cm$^2$As. $T = 4.2$ K. (b) Replotted as $d\rho_{xx}/dB$. (c) Numerically calculated density of states for $D_\perp = 0$ cm$^2$As. The arrows in panels (d) and (f) show schematic positions of the zero-mode Landau levels.

**Fig. 7** Comparison of Landau level spectra with dispersion relations for AB-stacked five-layer graphene.

(a) (Top) Numerically calculated Landau level spectra for $D_\perp = 1.72 \times 10^{-7}$ cm$^2$As. Numbers show filling factors of some energy gaps (Bottom) Numerically calculated dispersion relation for the same $D_\perp$. Some points in the dispersion relation are labeled $a$, $b_e$, $b_h$ and $m$. (b) Landau level spectra and dispersion for $D_\perp = 0$.

**Fig. 8** Landau fan diagram of AB-stacked six-layer graphene.
(a) Map of $\rho_{xx}$ as a function of $n_{tot}$ and $B$ measured for $D_\perp = 1.97 \times 10^{-7}$ cm$^{-2}$As. $T = 4.2$ K. (b) Map of $\rho_{xx}$ measured for $D_\perp = 0$ cm$^{-2}$As. $T = 4.2$ K. Arrows indicate schematic positions of the zero-mode Landau levels. (c) Map of numerically calculated density of states $D_\perp = 1.92 \times 10^{-7}$ cm$^{-2}$As using Slonczewski-Weiss-McClure parameters of graphite. (d) Numerically calculated density of states for $D_\perp =$

0 cm$^{-2}$As. Arrows indicate schematic positions of the zero-mode Landau levels.

Fig. 9 Landau level spectra and dispersion relation in AB-stacked six-layer graphene. (a) (Top) Numerically calculated Landau level spectra for $D_\perp = 1.92 \times 10^{-7}$ cm$^2$As. (bottom). Numerically calculated dispersion relation for the same $D_\perp$. Result for K point is shown. Characteristic points in the band are labeled $a - j$. (b) Structures associated mini-Dirac cones are indicated on the maps of the density of states (left) and $\rho_{xx}$(right) by A - C.

Fig. 10

Effect of trigonal warping on Landau spectra.

(a) Numerically calculated density of states of AB-stacked four-layer graphene by using $\gamma_3 = 0$ (left) and 0.3 eV (right). Other SWMcC parameters are the same as those of graphite. $D_\perp = 3.8 \times 10^{-7}$cm$^2$As. (b) Similar plots for AB-stacked five-layer graphene. $D_\perp = 1.72 \times 10^{-7}$ cm$^2$As. (c) Similar plots for AB stacked six-layer graphene. $D_\perp = 1.92 \times 10^{-7}$ cm$^2$As.

Fig. 11

Effect of trigonal warping on the Landau spectra in high magnetic fields.

(a) (Top) Numerically calculated density of states of AB-stacked six-layer graphene for $\gamma_3 = 0.3$ eV. (Bottom) Dispersion relation for K point. Other SWMcC parameters are the same as those of graphite. $D_\perp = 3.8 \times 10^{-7}$cm$^2$As. (b) Similar plot for $\gamma_3 = 0$ eV. Zero-mode Landau levels arising from the same the bilayer-like bands are grouped together in the ovals. Dashed lines indicate energies of the bottoms of the bilayer-like bands. $D_\perp$ is the same as that for panel (a).

# Fig1

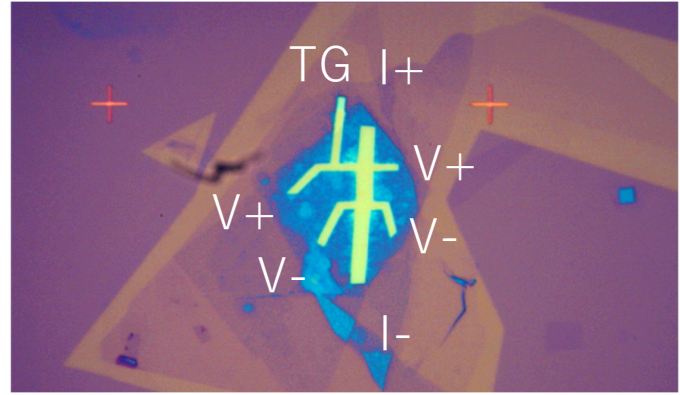
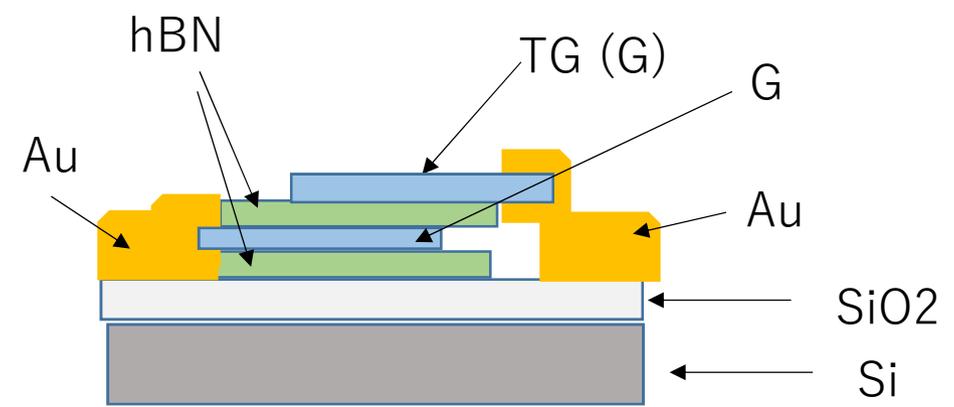
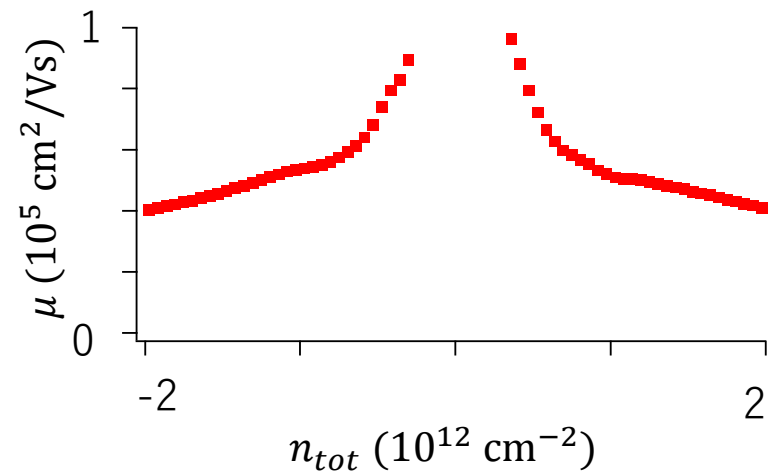

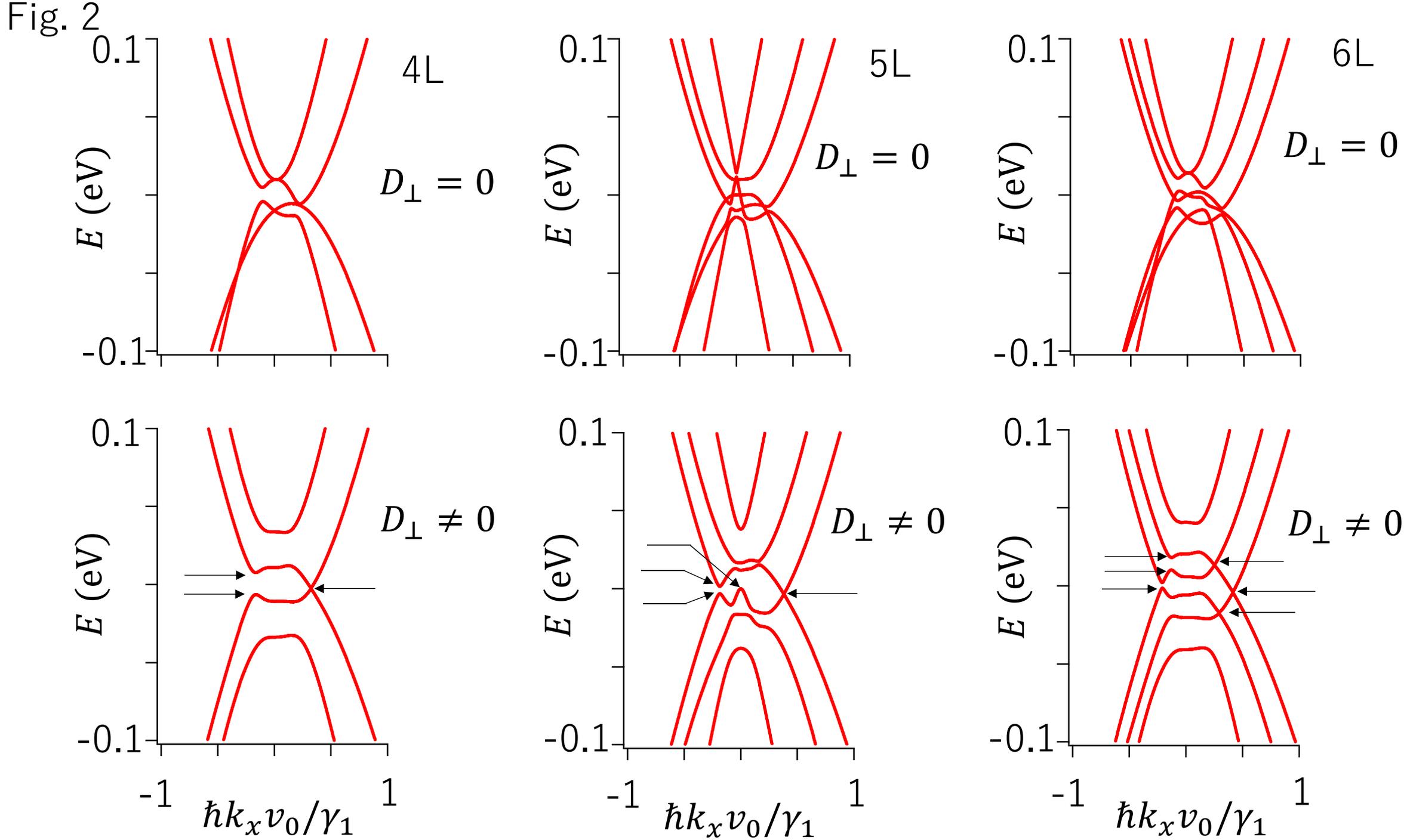

Fig. 2

Fig. 3

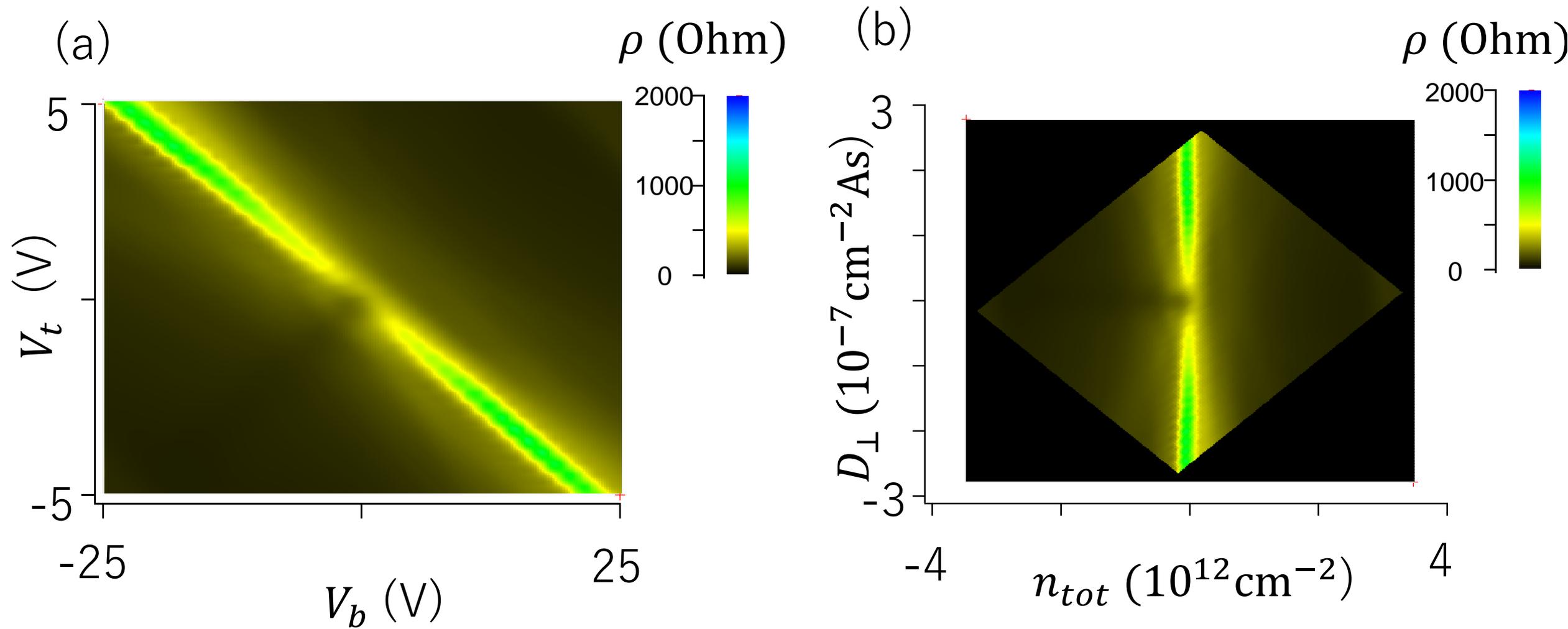

Fig4

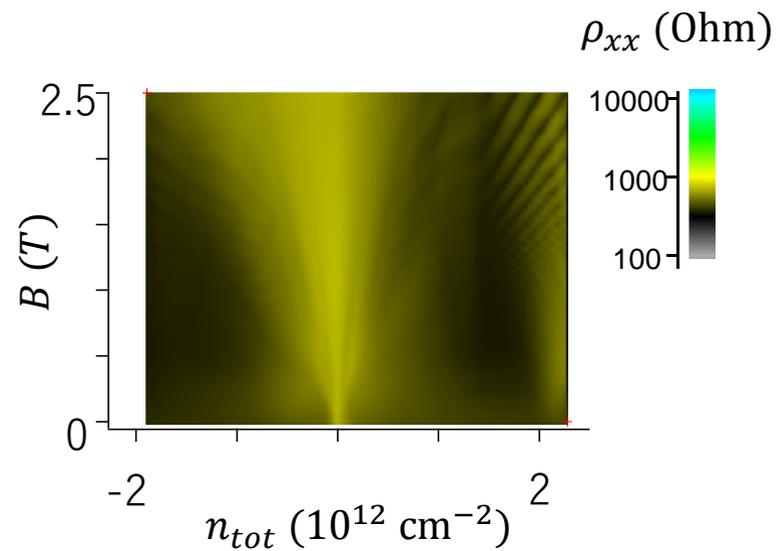
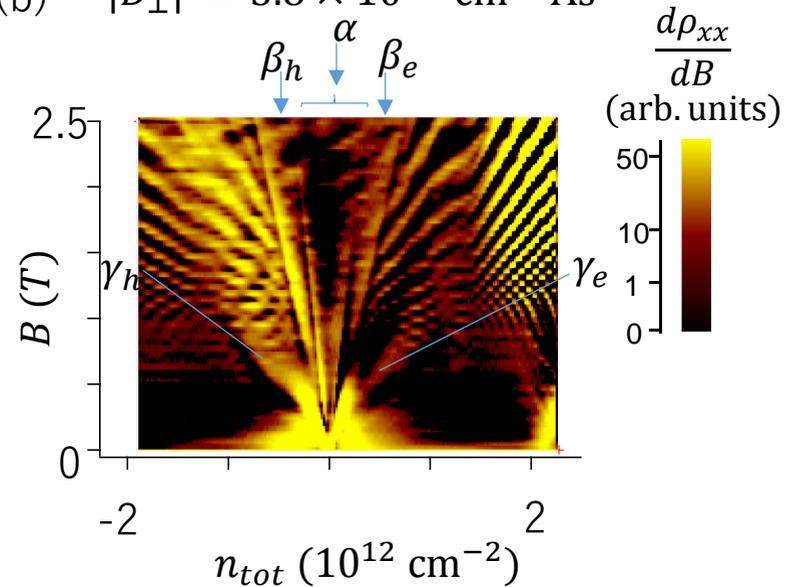
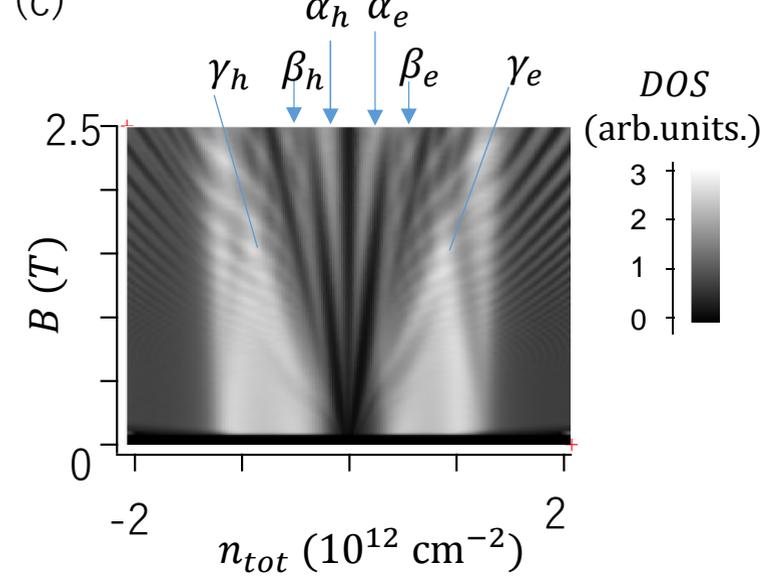
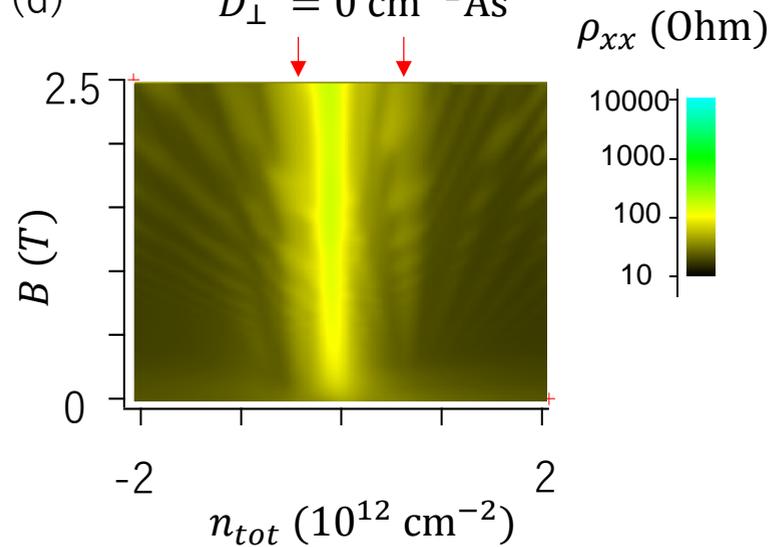
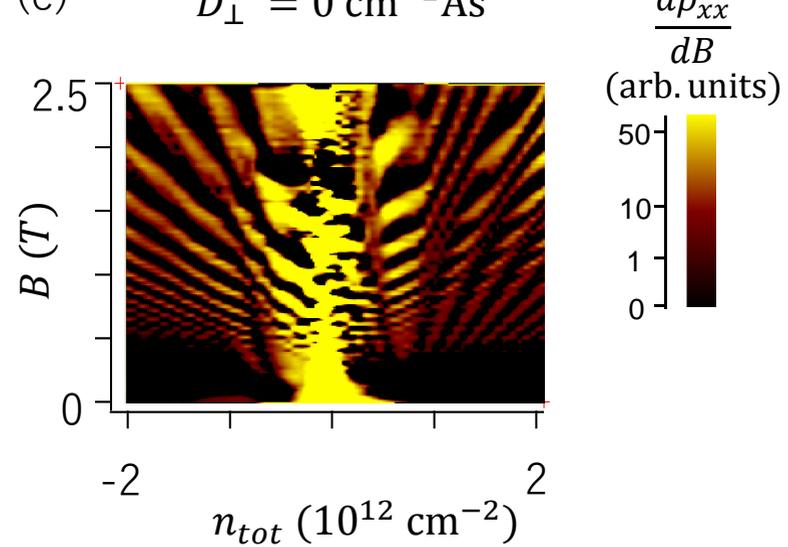
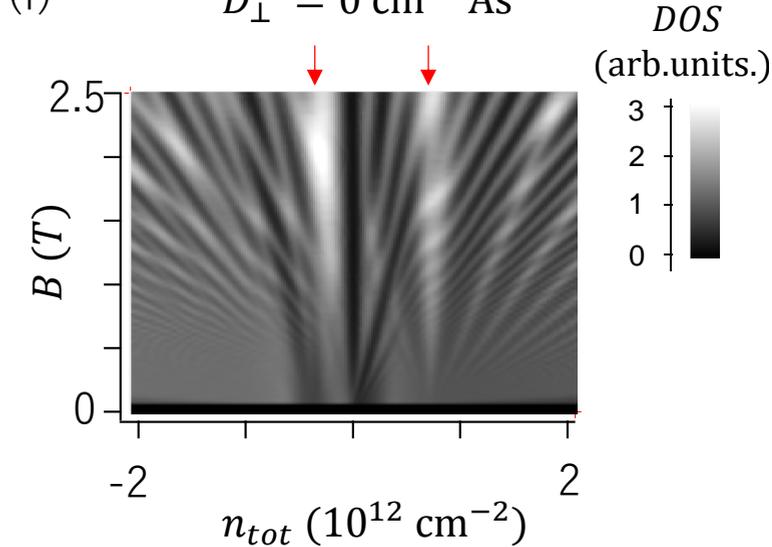

Fig5

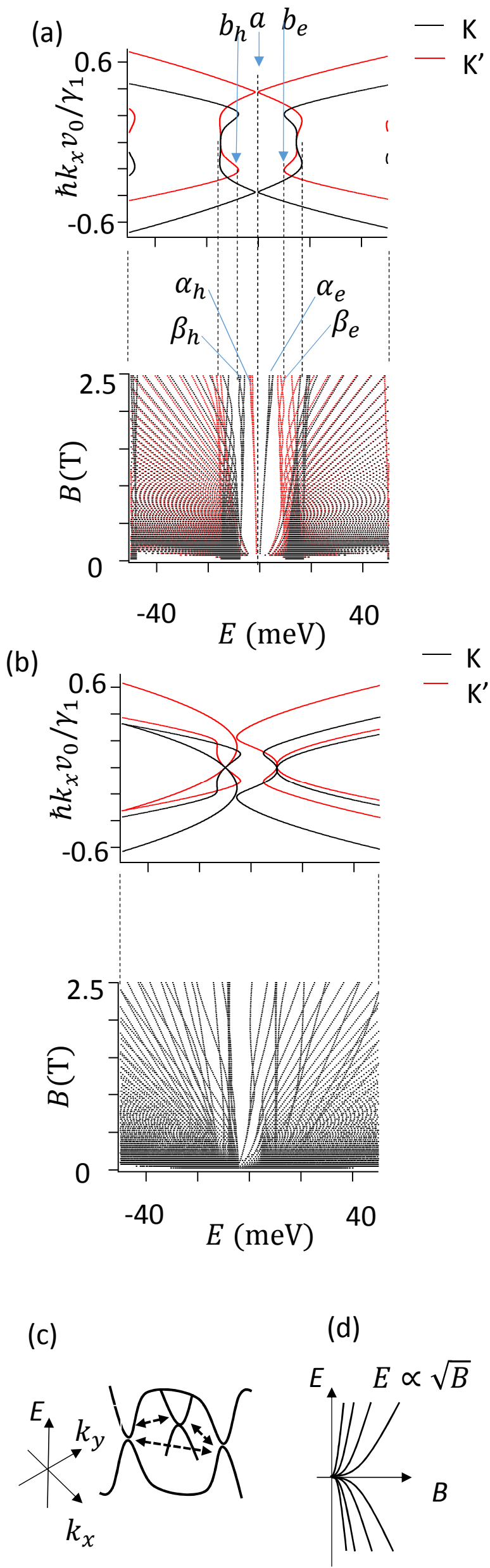

Fig6

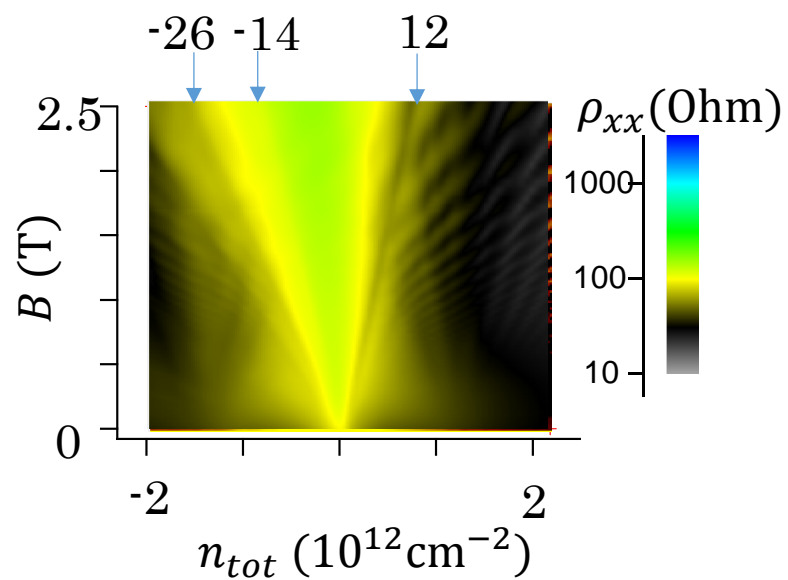 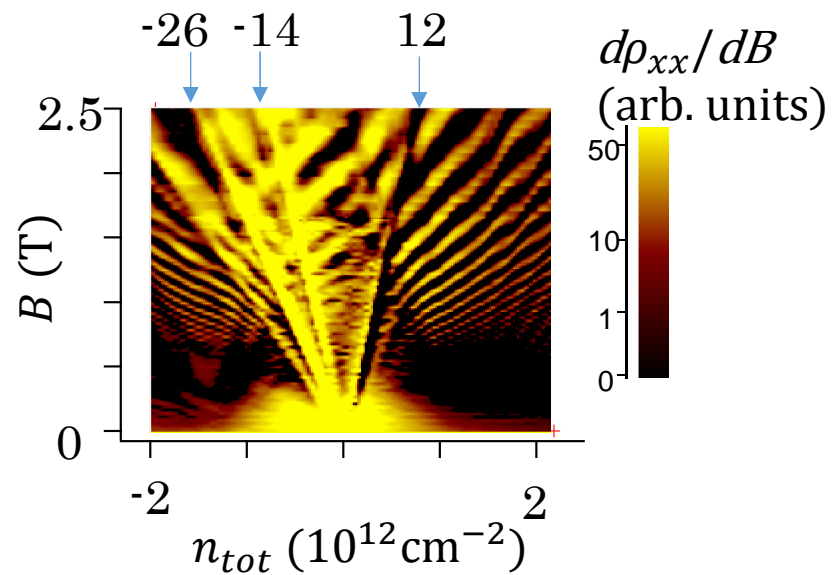 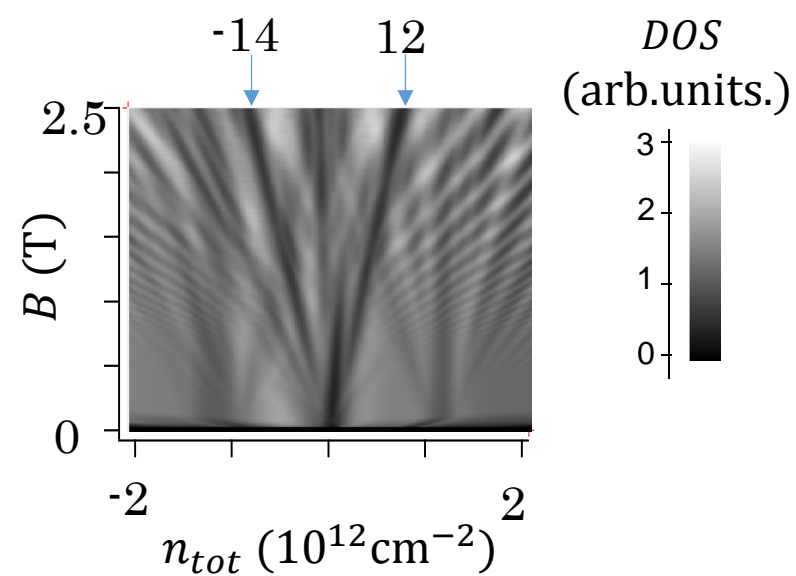

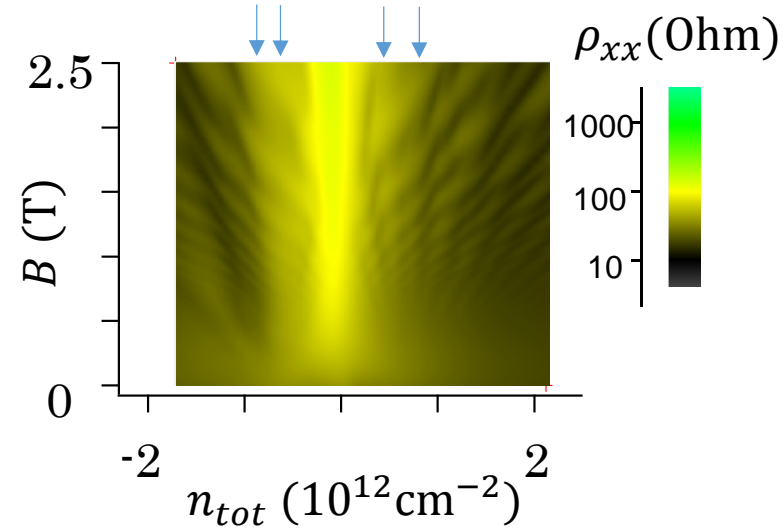 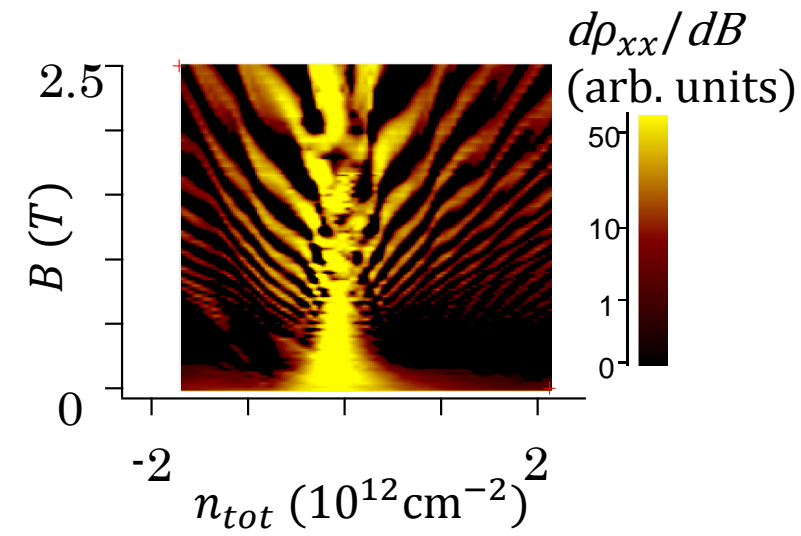 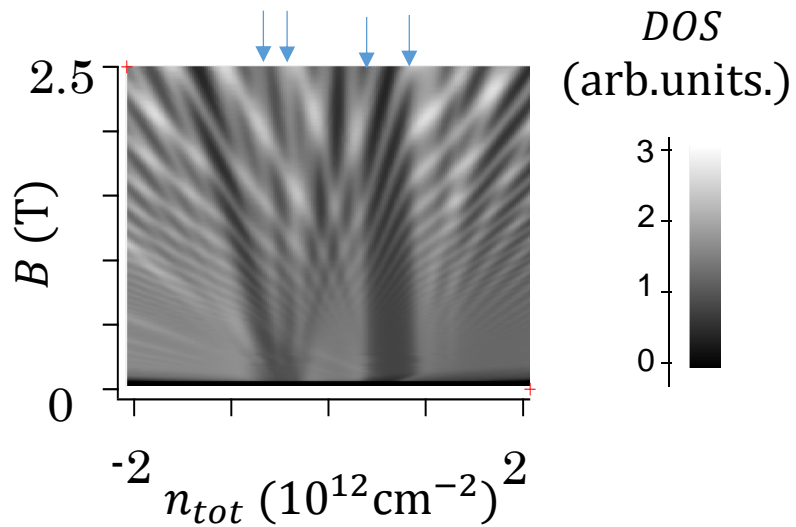

Fig7

(a) $D_\perp = 1.72 \times 10^{-7}$ cm$^{-2}$ As

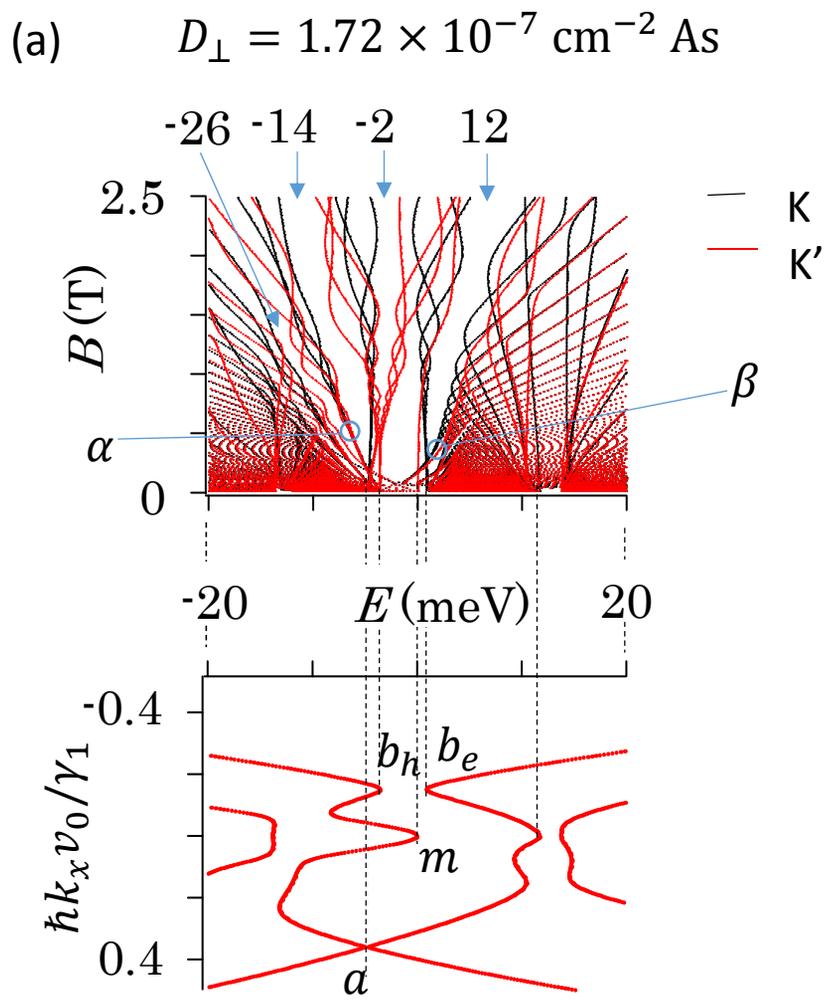

(b) $D_\perp = 0$ cm$^{-2}$ As

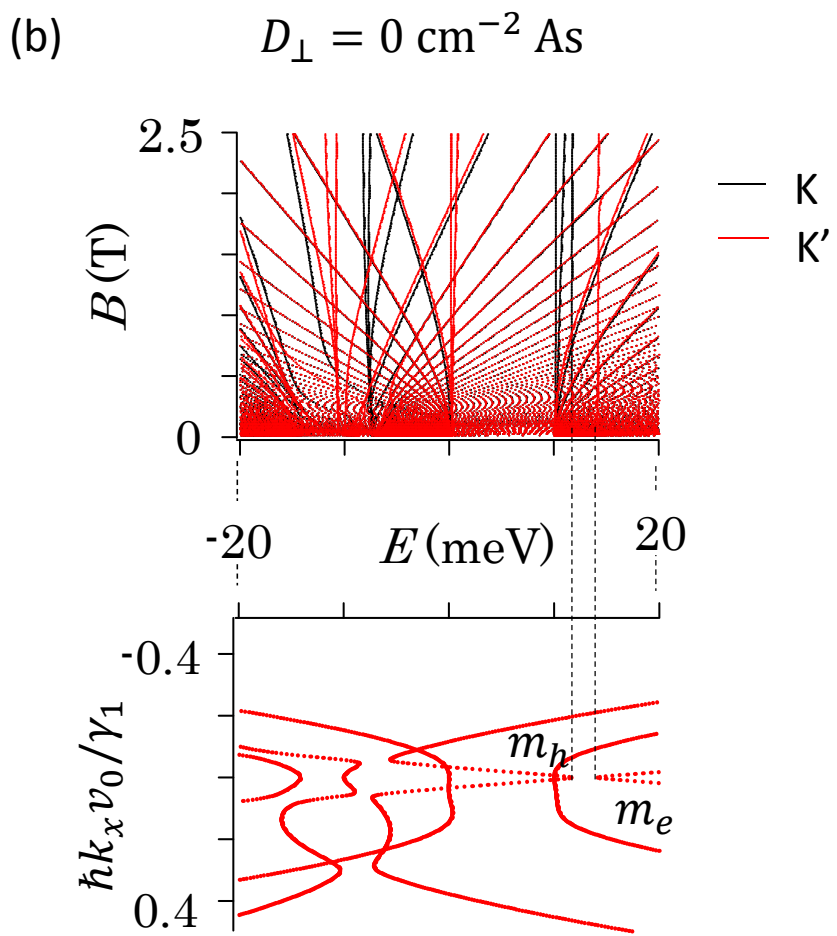

Fig 8

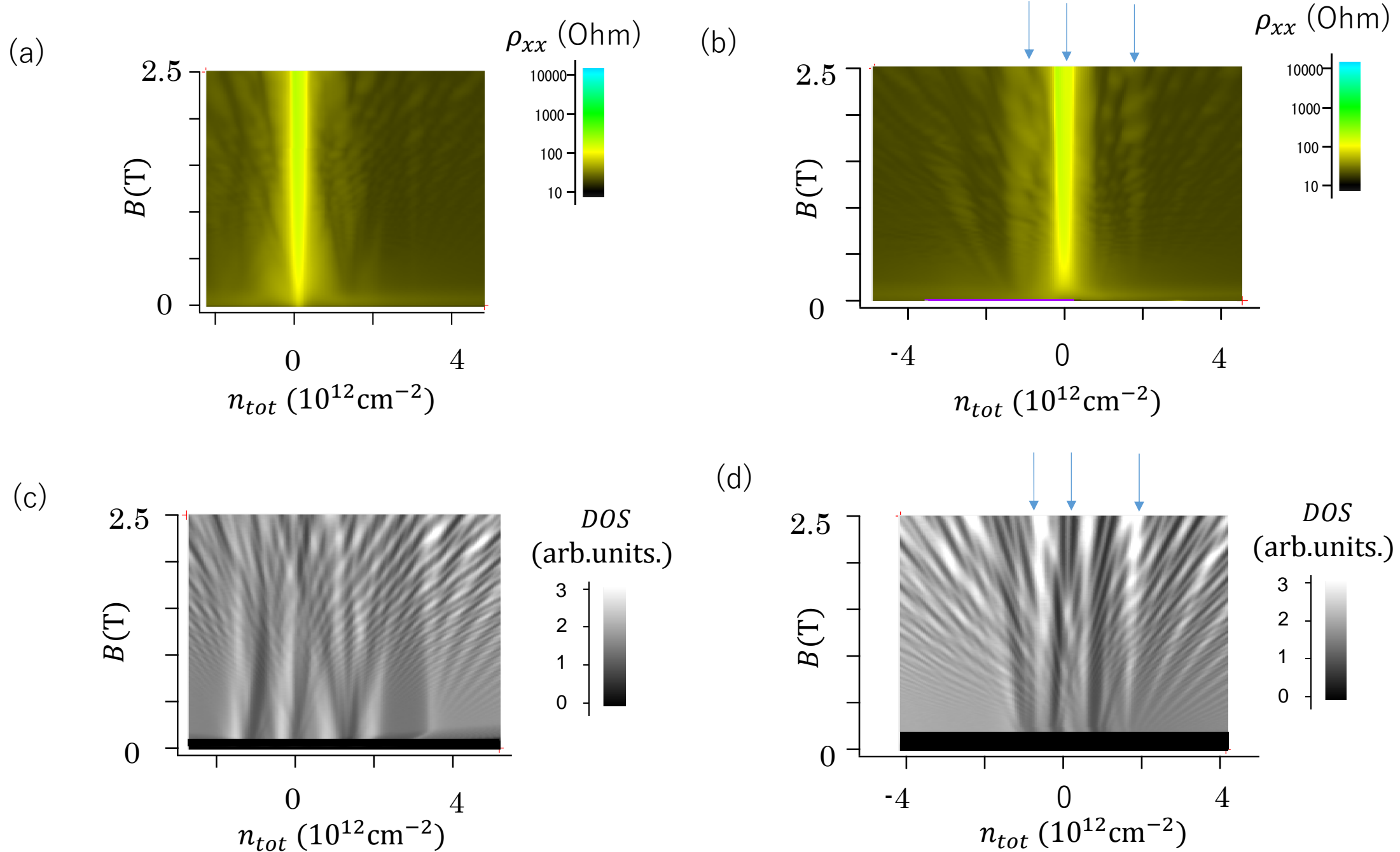

Fig 9

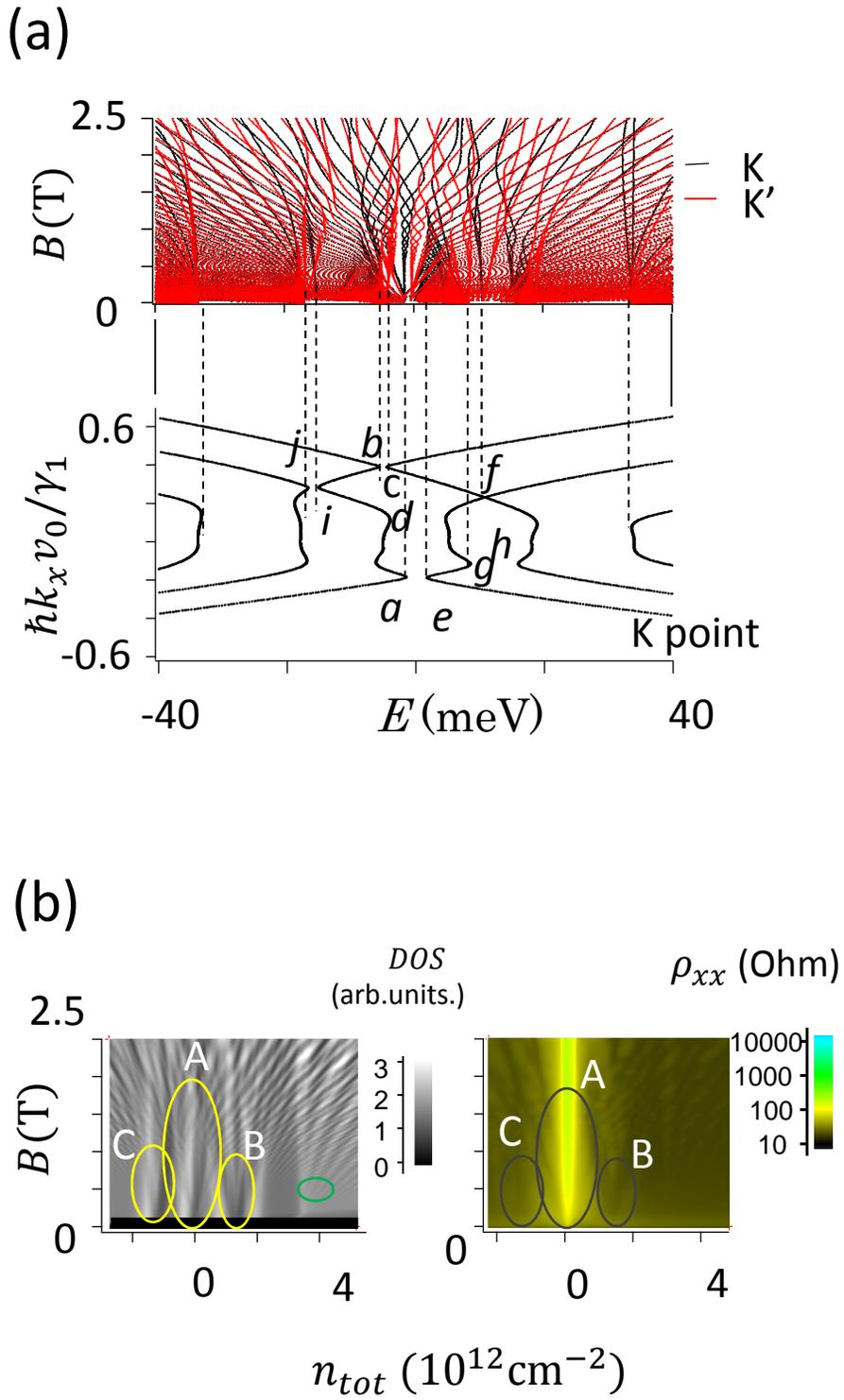

# Fig 10

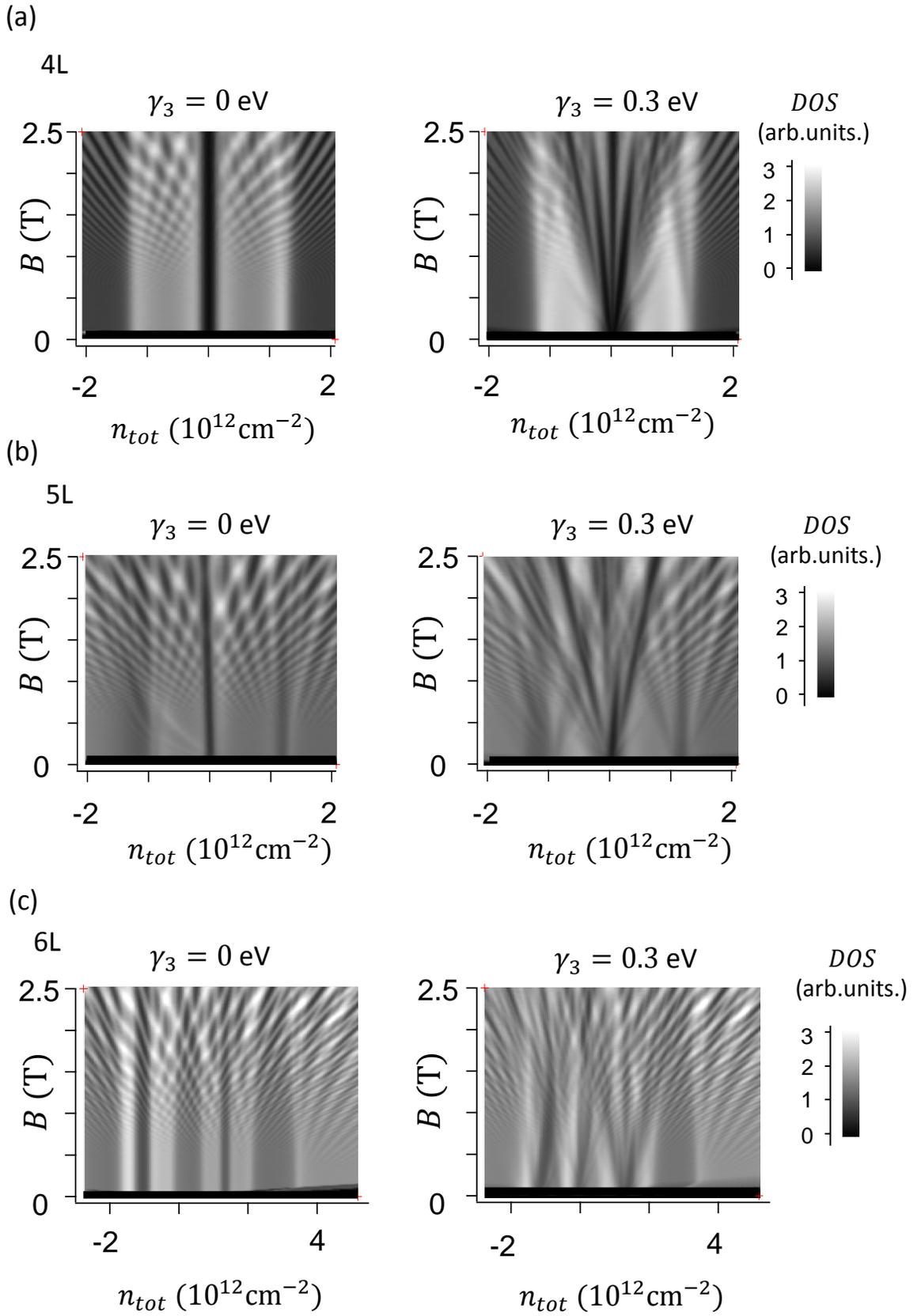

Fig 11

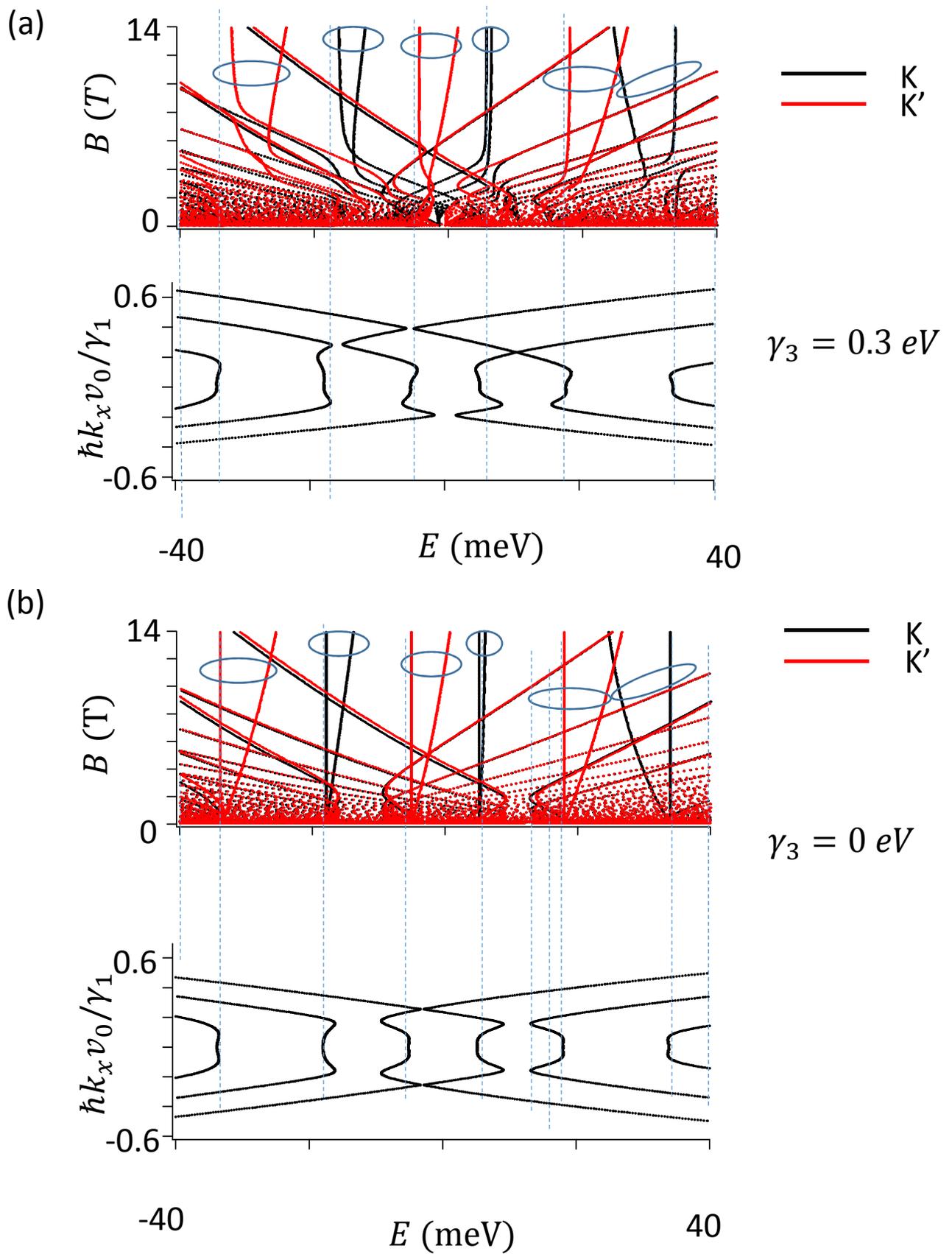